\documentclass[%
  hyphens
]{jacow}
\usepackage[english]{babel}
\begin{document}

\title{Advanced Externally Seeded FEL Schemes\\ for High-Repetition-Rate Operation at SHINE}

\author{H. Yang}
\author{N. Huang}
\author{Z. Qi}
\author{T. Liu\thanks{liut@sari.ac.cn}}
\author{H. Deng\thanks{denghx@sari.ac.cn}}
\author{B. Liu}
\affil{Shanghai Advanced Research Institute, Chinese Academy of Sciences, Shanghai, China}

\maketitle

\begin{abstract}
Externally seeded free-electron lasers (FELs) are promising approaches for generating fully coherent soft-X-ray radiation. Their extension to shorter wavelengths and MHz-level repetition rates is, however, constrained by the limited availability of high-repetition-rate seed lasers with sufficient energy modulation. Recent self-amplification and direct-amplification experiments at the Shanghai Soft X-ray FEL facility have significantly relaxed the peak-power requirement for high-gain harmonic generation (HGHG) and opened a practical path toward echo-enabled harmonic generation (EEHG). Using the SHINE bypass line, three compatible high-repetition-rate seeded-FEL configurations are explored:  self-modulation cascaded HGHG, self-modulation EEHG, and direct-amplification-driven EEHG. Numerical simulations indicate that these schemes can provide flexible routes toward MHz-level operation with harmonic generation beyond the 30th order. A common modulator-chicane layout is proposed to preserve compatibility among the candidate modes and to support future optimization and experimental implementation at SHINE.
\end{abstract}

\section{INTRODUCTION}
Externally seeded FELs provide a route to fully coherent extreme-ultraviolet and soft-X-ray pulses by imprinting a laser-induced energy modulation on a high-brightness electron beam and converting it into high-harmonic density bunching. High-gain harmonic generation (HGHG) is conceptually simple and has been widely applied, but its harmonic up-conversion efficiency usually requires a normalized energy modulation amplitude comparable to the desired harmonic number~\cite{Yu:PRA1991}. Echo-enabled harmonic generation (EEHG) can reach much higher harmonics, but it also imposes stringent requirements on seed lasers, electron-beam quality, and phase-space preservation~\cite{Stupakov:PRL2009}.

For high-repetition-rate FEL facilities, the main challenge is the limited pulse energy and peak power of ultraviolet seed lasers. This limitation becomes particularly severe for MHz-level operation, where the available seed pulse energy is much lower than that used in conventional seeded FELs. Recent experiments at the Shanghai Soft X-ray FEL facility (SXFEL) have demonstrated self-amplification of coherent energy modulation and direct-amplification enabled harmonic generation, indicating that weak-seed operation can significantly reduce the laser requirement while preserving high-harmonic bunching~\cite{Yan:PRL2021,Yang:APN2023,Yang:FR2026,Qi:PRL2025}. These results motivate the study of advanced seeded-FEL schemes for SHINE.

This paper summarizes\vbox to6mm{} the design study of high-repetition-rate externally seeded FEL options for the SHINE FEL-II with the bypass line. The goal is to identify flexible configurations that can support medium-high repetition-rate operation with standard EEHG and can be upgraded toward MHz-level operation using weak-seed enhancement schemes. The emphasis is placed on layout compatibility, achievable harmonic range, and experimental maturity.

\section{SHINE BYPASS LINE and\\ FEL-II PARAMETERS}
The SHINE bypass line provides a lower-energy electron beam of 3.0-4.5 GeV that is advantageous for laser-induced modulation and for extending the FEL-II photon-energy coverage toward the soft-X-ray range~\cite{Liu:FrontPhys2023}. The representative parameters used in the present optimization are listed in Table~\ref{tab:beam}. The lower beam energy facilitates energy modulation by a weak ultraviolet seed laser, while the bypass configuration provides an independent platform for externally seeded FEL development. The present study uses idealized beam distributions to define nominal working points and to compare different seeded configurations.

\begin{table}[!h]
\centering\vskip2mm
\caption{Representative Beam and Modulator Parameters Used in the Seeded-FEL Optimization}\vskip-1.5mm
\label{tab:beam}
\begin{tabular}{ll}
\toprule
\bf Parameter & \bf Value \\
\midrule
Beam energy & \qty{4}{GeV} \\
Peak current & $\sim$\qty{800}{A} \\
Bunch charge & \qty{100}{pC} \\
Bunch length (FWHM) & $\sim$\qty{120}{fs} \\
Normalized slice emittance & \qty{0.5}{\milli\metre\milli\radian} \\
Slice energy spread & $\sim$\qty{400}{keV} \\
Modulator length & 3/6/\qty{9}{m} \\
Modulator period & \qty{140}{mm} \\
\bottomrule
\end{tabular}
\end{table}

\section{HIGH-REPETITION-RATE SEEDED\\ FEL OPTIONS}
\begin{figure}[t]
\centering
\includegraphics[width=\linewidth]{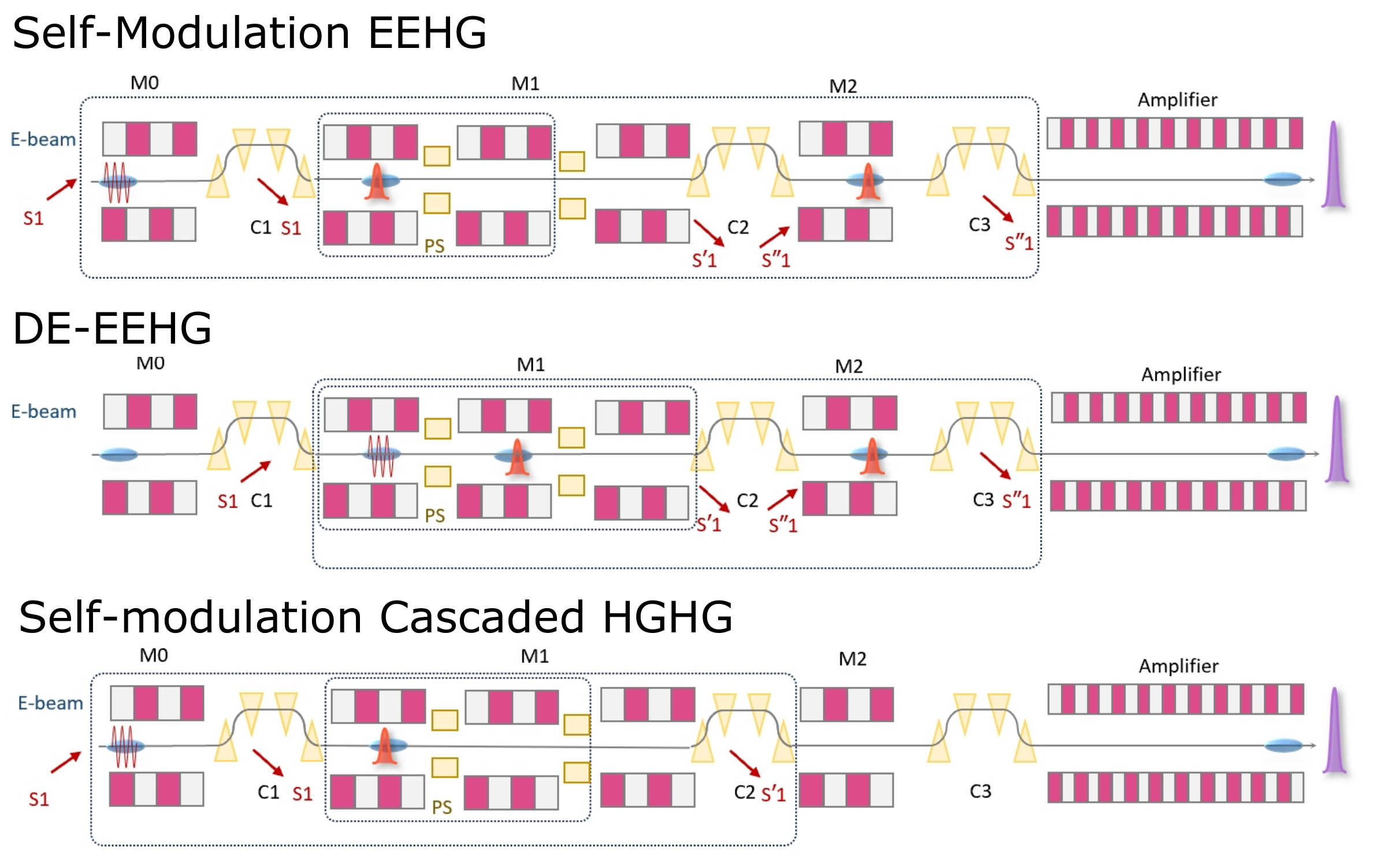}\vskip-1mm
\caption{Conceptual comparison of three MHz-level externally seeded FEL options for SHINE.}
\label{fig:schemes}\vskip2mm
\end{figure}

Three weak-seed schemes are considered for MHz-level operation, as schematically summarized in Fig.~\ref{fig:schemes}. They reuse a common modulator-chicane infrastructure while providing different levels of harmonic reach, layout simplicity, and experimental maturity. In the self-modulation-based options, the self-modulator can also be tuned to a harmonic resonance of the seed. This allows harmonic bunching to be produced at shorter effective wavelengths, useful for extending the output wavelength range toward higher\vtop to6mm{} harmonics.

\subsection{Self-Modulation EEHG}
The self-modulation EEHG (SM-EEHG) scheme combines self-amplified coherent radiation with EEHG-like phase-space manipulation. A weak seed first generates an initial modulation, the following self-modulator amplifies the coherent radiation, and the delayed radiation is refocused to drive the second energy modulation. The self-modulator may be resonant at a harmonic of the seed, enabling shorter-wavelength coherent radiation and enhanced high-harmonic bunching.

As a representative working point, a \qty{10}{MW} seed laser is used. The modulation amplitudes and dispersive strengths are optimized to produce a pronounced high-bunching region, as shown in Fig.~\ref{fig:results}(a). The bunching distribution indicates that the selected working point lies within a reasonably localized optimum in the $R^1_{56}$--$R^2_{56}$ parameter space. Subsequent FEL simulations show GW-level peak power, few-femtosecond pulse duration, and tens-of-microjoule pulse energy, as illustrated by the spectrum in Fig.~\ref{fig:results}(b). SM-EEHG is therefore considered a promising MHz-level candidate, although direct experimental validation is still required.

\subsection{Direct-Amplification-Driven EEHG}
In the direct-amplification-driven EEHG (DE-EEHG) scheme, coherent radiation induced by a weak seed laser is directly amplified in a long modulator and then reused as the effective second seed in an EEHG configuration. The layout is relatively simple and remains highly compatible with the standard EEHG baseline. Recent direct-amplification enabled harmonic-generation experiments at SXFEL have demonstrated the relevant principle, including operation with a weak seed laser and stable harmonic output~\cite{Qi:PRL2025}.

For a typical optimized case, a \qty{20}{MW} seed laser is adopted, and the modulation amplitudes and chicane strengths are adjusted to generate strong high-harmonic bunching. As shown in Fig.~\ref{fig:results}(c), a clear bunching maximum is obtained around the selected working point, supporting efficient harmonic conversion at the 30th harmonic. The corresponding FEL spectrum in Fig.~\ref{fig:results}(d) shows narrow-band coherent output. Owing to its simple layout and good compatibility with the baseline EEHG configuration, DE-EEHG is regarded as a practical MHz option, particularly for the 20th--40th harmonic range.

\subsection{Self-Modulation Cascaded HGHG}
In the self-modulation cascaded HGHG scheme, a single weak ultraviolet seed generates a small initial energy modulation, which is subsequently enhanced by coherent radiation in a self-modulation section. The self-modulator can be tuned to a seed harmonic, allowing the first stage to generate shorter-wavelength coherent radiation and thereby improving the harmonic reach of the cascaded HGHG process. The relevant self-modulation mechanism has been experimentally demonstrated at SXFEL through self-amplification and high-harmonic lasing~\cite{Yan:PRL2021,Yang:APN2023}.

\begin{figure}[t]
\centering
\includegraphics[width=0.45\textwidth]{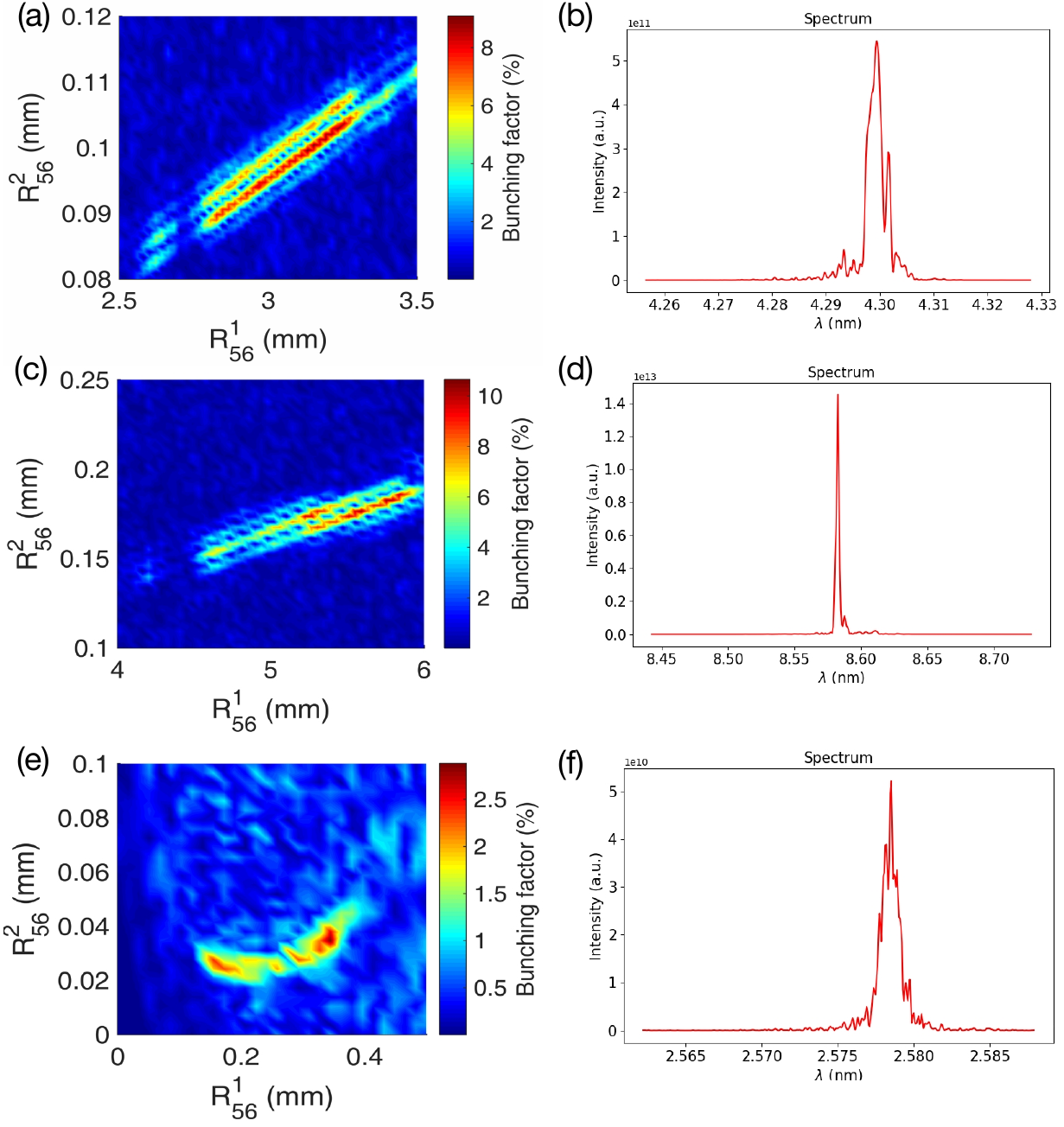}\vskip-1.5mm
\caption{Representative optimization and FEL output for the three MHz-level seeded-FEL options. Panels (a) and (b) show the SM-EEHG bunching scan at the 60th harmonic and the corresponding FEL spectrum. Panels (c) and (d) show the DE-EEHG bunching scan at the 30th harmonic and the FEL spectrum. Panels (e) and (f) show the self-modulation cascaded HGHG optimization and final spectrum.}
\label{fig:results}\vskip2mm
\end{figure}

For the representative SHINE case shown in Fig.~\ref{fig:results}(e,f), a single \qty{20}{MW} seed pulse, corresponding to sub-microjoule pulse energy, is used. The first stage generates GW-level coherent radiation with a femtosecond-scale pulse duration, which is then used to seed the following HGHG stage. The final FEL output reaches the microjoule level with a narrow relative bandwidth and a near-Fourier-limited time-bandwidth product. This option provides a flexible alternative route toward high harmonics at MHz repetition rate, but the induced energy spread from the first stage should be carefully controlled.

Table~\ref{tab:schemes} summarizes the main role and experimental status of the candidate modes. Standard EEHG and EEHG-HGHG cascade are considered near-term baseline options for 10--\qty{100}{kHz} operation. For MHz-level operation, the final choice should be deferred until reliable operational data on the linac and FEL become available, especially the measured slice energy spread and timing stability.

\begin{table}[!h]
\centering\vskip1mm
\caption{Candidate Externally Seeded FEL Options}\vskip-1mm
\label{tab:schemes}
\small
\begin{tabular}{@{}lccc@{}}
\toprule
Mode & Harm. & Rep. & Role \\
\midrule
EEHG & 50--100 & 10--100 kHz & Base \\
EEHG-HGHG cascade & 50--250 & 10--100 kHz & Upg. \\
SM-EEHG & 20--80 & 1 MHz & Sim. \\
DE-EEHG & 20--40 & 1 MHz & POP \\
SM-HGHG cascade & 20--80 & 1 MHz & POP \\
\bottomrule
\end{tabular}
\end{table}

\begin{figure}[b]
	\centering\vskip3mm
	\includegraphics[width=0.8\linewidth]{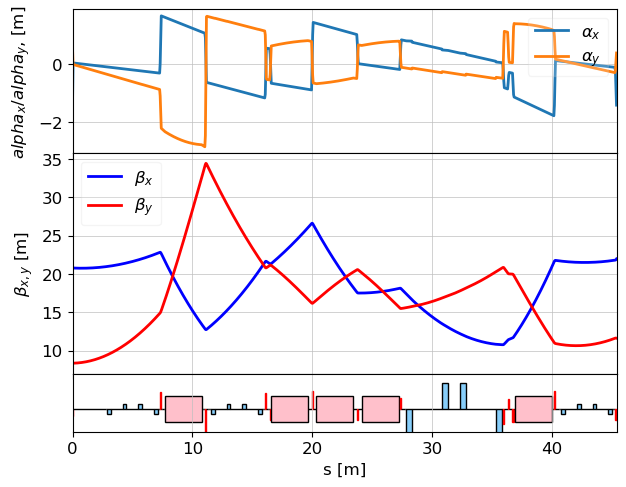}\vskip-1.5mm
	\caption{Lattice design and matching result for the modulator-chicane section.}
	\label{fig:lattice}
\end{figure}

\section{KEY PRACTICAL CONSIDERATIONS}
A shared hardware platform is proposed to preserve flexibility among the candidate seeded FEL options. The baseline layout consists of five 3-m-long modulators with a period of \qty{140}{mm}, together with four chicanes, including one large chicane and three small chicanes. Three seed-laser injection points and one optical delay platform are foreseen. This arrangement can support standard EEHG, EEHG-HGHG cascade, SM-HGHG cascade, SM-EEHG, and DE-EEHG with limited changes to the magnetic layout.

The optics matching of the modulator section is shown in Fig.~\ref{fig:lattice}. The matched lattice provides controlled beta functions and transverse beam sizes throughout the modulator-chicane section, supporting stable beam transport and compatibility with the different seeded-FEL configurations. The lattice provides a practical compromise between optical access, chicane installation, and beta-function control. Weak focusing from the modulator modules is included in the matching and will be further refined together with the detailed magnetic and mechanical design.

\section{CONCLUSION and discussion}
A set of externally seeded FEL options has been studied for high-repetition-rate operation at SHINE. Standard EEHG and EEHG-HGHG cascade provide near-term baseline routes for 10--\qty{100}{kHz} operation, while SM-EEHG, DE-EEHG, and SM-HGHG cascade provide flexible candidates for MHz-level operation with a single weak seed laser. A common modulator-chicane layout is proposed to preserve compatibility among these modes. SM-HGHG cascade and DE-EEHG are supported by relevant proof-of-principle experiments at SXFEL, whereas SM-EEHG remains a promising candidate requiring further experimental demonstration.

A systematic sensitivity and jitter study will be performed to define realistic tolerances, robust working points, and optimization targets for the seeded FEL schemes. The analysis including slice energy spread, peak current, emittance, residual energy chirp, current-profile variation, timing jitter, and seed-laser energy, collective effects, and microbunching instability will be presented in a dedicated journal paper. SXFEL will continue to serve as an important experimental testbed for weak-seed modulation, coherent radiation amplification, optical synchronization, and high-harmonic generation to reduce the technical risk for SHINE implementation.

\section{ACKNOWLEDGEMENTS}
This work was supported by the National Key Research and Development Program of China (2024YFA1612101), the National Natural Science Foundation of China (12125508, 12541503), the Shanghai Pilot Program for Basic Research – Chinese Academy of Sciences, Shanghai Branch (JCYJ-SHFY-2021-010), the Innovation Program of Shanghai Advanced Research Institute, CAS (2025CP006), and the China Postdoctoral Science Foundation (2025M770914).

\end{document}